\documentclass[
  final            
,a4paper
  ]{aipproc}

\usepackage{natbib}

\layoutstyle{6x9}

\newcommand{\g}{$\gamma$}
\newcommand{\wco}{W_\mathrm{CO}}
\newcommand{\xco}{X_\mathrm{CO}}
\newcommand{\nhd}{N(\mathrm{H_2})}
\newcommand {\lesssim} {\ {\raise-.5ex\hbox{$\buildrel<\over\sim$}}\ }
\newcommand{\hi}{\mathrm{H\,\scriptstyle{I}}}
\newcommand{\nhi}{N(\mathrm{H\,\scriptstyle{I}})}
\newcommand{\hd}{\mathrm{H}_2}

\newcommand{\apj}{Astrophys. J. }
\newcommand{\apjl}{Astrophys. J. Lett. }
\newcommand{\apjs}{Astrophys. J. Suppl. }
\newcommand{\aap}{Astron. \& Astrophys. }
\newcommand{\astropp}{Astroparticle Phys. }

\newcommand{\prl}{Phys. Rev. Lett. }
\newcommand{\pasj}{Publ. Astron. Soc. Jpn. }

\begin{document}

\title[Diffuse \g-ray emission with \textit{Fermi} LAT]{\textit{Fermi}
LAT measurements
of
diffuse \g-ray emission: results at the first-year milestone}


\author{Luigi Tibaldo, on behalf of the \textit{Fermi} LAT collaboration}{
  address={INFN -- Sezione di Padova and Dipartimento di Fisica ``G. Galilei''
-- Universit\`a di Padova, I-35131 Padova, Italy},
  altaddress={Laboratoire AIM, CEA-IRFU/CNRS/Universit\'e Paris Diderot, Service
d'Astrophysique, CEA Saclay, 91191 Gif sur Yvette, France}
}

\begin{abstract}

For more than one year the \textit{Fermi} Large Area Telescope has been
surveying the \g-ray sky from 20 MeV to more than 300 GeV with
unprecedented statistics and angular resolution. One of the key science
targets of the \textit{Fermi} mission is diffuse \g-ray emission.
Galactic interstellar \g-ray emission is produced by interactions of
high-energy cosmic rays with the interstellar gas and radiation field. We
review the most important results on the subject obtained so far: the
non-confirmation of the excess of diffuse GeV emission seen by EGRET, the
measurement of the \g-ray emissivity spectrum of local interstellar gas, the
study of the gradient of cosmic-ray densities and of the $\xco=\nhd/\wco$ ratio
in the outer Galaxy. We also catch a glimpse at diffuse \g-ray emission in the
Large Magellanic Cloud. These results allow the improvement of large-scale
models of Galactic diffuse \g-ray emission and new measurements of the
extragalactic \g-ray background.

\end{abstract}

\maketitle


\section{Introduction}

Since the dawn of \g-ray astronomy the \g-ray sky above a few tens MeV has been
known to be dominated by diffuse emission \citep{Clark1968}, giving more than
80\% of the observed photons:
\begin{itemize}
 \item a bright component is correlated with the Milky Way structures, and
thus is interpreted to be Galactic in origin, arising from interactions of
high-energy cosmic rays (CRs) with the gas in the interstellar medium (ISM) and
the interstellar radiation field (ISRF);
 \item a weaker component is observed with almost isotropic distribution
over the sky, and thus is thought to be extragalactic in origin and
usually referred to as the extragalactic \g-ray background (EGB). 
\end{itemize}
Unresolved point sources also contribute to what we call \emph{diffuse}
emission, remarkably for the EGB, which might be, according to current
estimates,
made up in large part by populations of unresolved extragalactic \g-ray sources
like blazars and external galaxies \citep{Dermer2007}.

Galactic interstellar \g-ray emission is produced in different interaction
processes:
\begin{itemize}
 \item CR nucleons interact with gas nuclei in the ISM
leading to \g-ray production through $\pi^0$ production and decay;
 \item CR leptons interact with gas nuclei in the ISM producing \g-rays via
Bremsstrahlung;
 \item CR leptons interact with low-energy photons of the ISRF producing
\g-rays via inverse Compton (IC) scattering.
\end{itemize}
Therefore, the interstellar \g-ray emission is a tracer of CR densities in the
Galaxy,
suitable to study CRs in distant locations that we cannot access with direct
measurements, as well as of the total ISM column densities, complementary to
studies at other wavelengths. Additionally, modeling the Galactic diffuse
emission is
fundamental for \g-ray astrophysics, since it provides a bright and structured
foreground for source detection and characterization, the study of the EGB and
the search for signals from exotic processes,
like annihilation of dark matter (DM) particles. Our
models of the interstellar emission result from the combination of a wide range
of
available information, including CR spectra and composition at Earth,
theoretical description of the propagation processes, multiwavelength studies
of the ISM/ISRF, measurements of the relevant interaction processes and
estimates
of the Galactic magnetic field.

Several studies on the EGB have been performed
\citep{Fichtel1977,Sreekumar1998}, but its nature is still
mysterious.
Apart from the contribution by unresolved sources mentioned above, many
processes which might produce \emph{truly diffuse} extragalactic emission have
been proposed \citep{Dermer2007}, for example large-scale
structure formation, interactions of ultra-high-energy CRs with the
extragalactic low-energy background radiation, annihilation or decay of
cosmological dark matter. On the other hand, interactions of CRs with
nearby matter, like solar system bodies \citep{Moskalenko2008} or debris at its
outer frontier \citep{Moskalenko2009}, might contribute to the EGB derived in
the aforementioned studies. Therefore, the extragalactic origin of this diffuse
component is still not clear, even though we keep referring to it as EGB.

A wealth of new information on high-energy diffuse \g-ray emission has been
provided during the last year by the Large Area Telescope (LAT) on board the
\textit{Fermi} observatory \citep{Atwood2009,Raino2009}, thanks to its large
acceptance, more than one order of magnitude larger than the predecessor EGRET,
and the improved angular resolution (e.g. at 1 GeV the single event 68\%
containment reaches
$\sim 0.6^\circ$ for the LAT compared with $\sim 1.7^\circ$ for EGRET). In this
contribution we review the most important results obtained in the first year of
LAT science operations,
which shed light on many open questions of the EGRET era.

\section{The EGRET GeV excess and LAT measurements at intermediate
Galactic latitudes}

One of the most intriguing legacies of the EGRET era was the so-called GeV
excess. Data by EGRET showed an excess of diffuse emission with respect to
conventional models based on the locally measured CR spectra, first reported
on the Galactic plane by \citet{Hunter1997}. It was then confirmed as
a $\sim 50\%$ excess over all directions in the sky, which led to a number of
possible interpretations, including instrumental effects, discrepancies between
locally measured CR spectra and local interstellar space and possible signals
of dark matter annihilation
\citep{Hunter1997,Strong2004EG,deBoer2005,stecker2008}.

In order to investigate this issue, the first target of LAT studies has been the
emission at intermediate Galactic latitudes, $10^\circ \leq |b| \leq
20^\circ$, where most of the emission is thought to be produced by interactions
of CRs with local interstellar matter ($\lesssim 1$ kpc from the Sun).
This region is well suited to verify if the observed \g-ray emission is
consistent with locally measured CR spectra.

\begin{figure}[!b]
\includegraphics[width=0.65\textwidth]{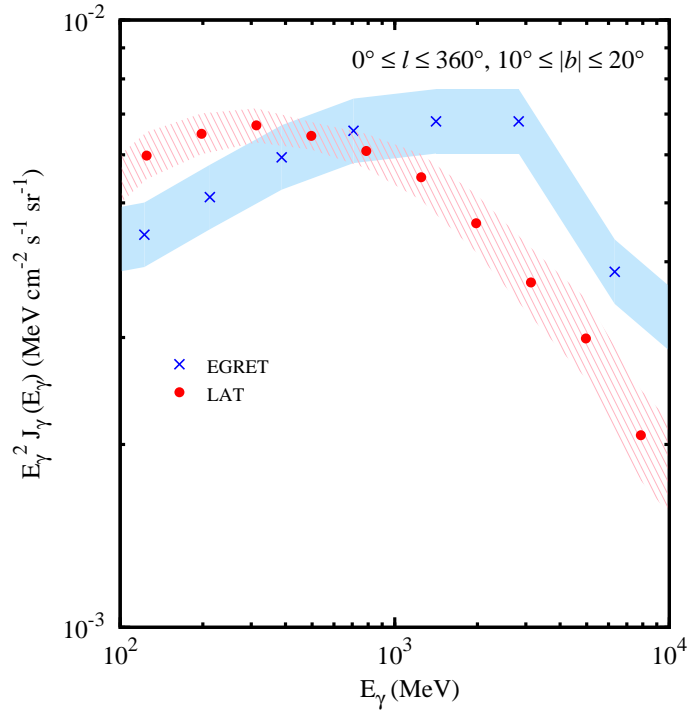}
\caption{\citep{noGeVexc} Spectrum of the $\gamma$-ray emission
measured by the LAT for energies between 100 MeV and 10 GeV, averaged over all
longitudes in the region with Galactic latitude $10^\circ \leq |b| \leq
20^\circ$ (circles), compared with the same spectrum as measured by EGRET
(crosses). Shaded bands represent the systematic errors, which are the
dominant source of uncertainty for both instruments.} \label{ngefig1}
\end{figure}

\begin{figure}[!t]
\includegraphics[width=0.65\textwidth]{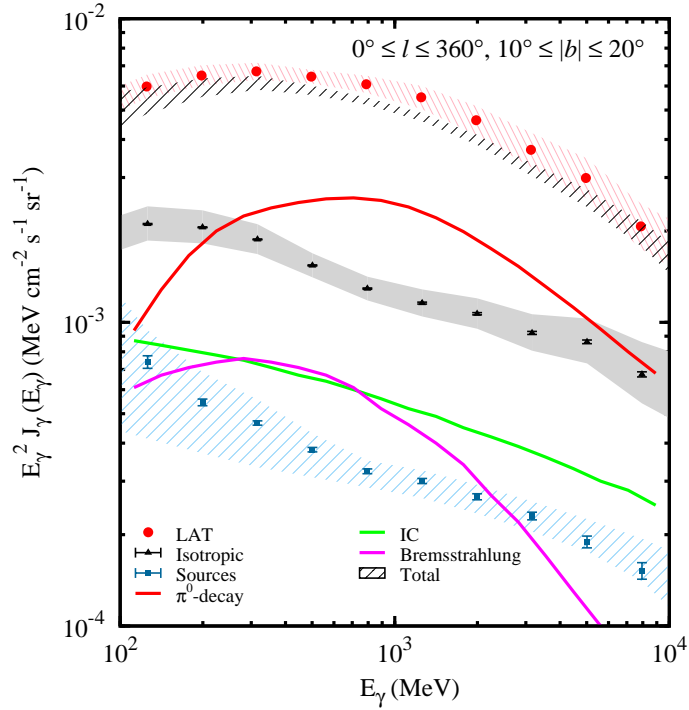}
\caption{\citep{noGeVexc} The spectrum by the LAT, as
already shown in Fig.~\ref{ngefig1}, is compared with the \emph{a priori}
GALPROP model, plus the contribution by resolved point sources and the
unidentified isotropic component (see text).} \label{ngefig2}
\end{figure}

LAT data have been compared with physical expectations using GALPROP, a code for
CR propagation in the Galaxy by
\citet{strong98,Strong2007}. The GALPROP model used for
this work is
a revised version of the ``conventional'' model described in
\citet{Strong2004EG}, based on the locally measured CR spectra, including
updated formalism to treat $p$-$p$ interactions, recent radio surveys of the
ISM, improved routines for line-of-sight integration and a complete
recalculation of the ISRF \citep[for details see][]{noGeVexc}. The results are
shown in Fig.~\ref{ngefig1} and~\ref{ngefig2}.

The LAT spectrum is softer than the EGRET one, showing significant lower
intensities for energies $>1$ GeV (Fig.~\ref{ngefig1}).
The LAT spectrum is
approximately reproduced by the GALPROP model of the Galactic interstellar
emission, plus the contribution from point
sources in the LAT 3-month source list and an unidentified isotropic component
derived from fitting LAT data with the same Galactic model fixed
(Fig.~\ref{ngefig2}).
LAT measurements at intermediate Galactic latitudes are thus inconsistent with
the
EGRET GeV excess and suggest that the local diffuse emission is explicable in
terms of interactions with interstellar matter of CRs with spectra similar to
those measured at Earth.

\section{The emissivity of local interstellar atomic hydrogen}\label{hiemisspar}
To quantitatively compare the local diffuse emission with \emph{in situ} CR
spectra, we studied the
emissivity spectrum of local interstellar atomic hydrogen, $\hi$,
based on measurements of the column densities $\nhi$ across the sky
derived from radio surveys of its 21 cm line.

For this purpose we selected a region of the sky at intermediate Galactic
latitudes, at $200 ^\circ \leq l \leq 260 ^\circ$ and $22 ^\circ \leq |b| \leq
60 ^\circ$, where most of the gas along the line of sight is nearby (once again
$\lesssim 1$ kpc from the solar system) and no known
complexes of molecular gas are present. The \g-ray intensity
maps, after subtraction of the components due to point sources and IC
emission, were correlated with
the distribution of $\nhi$, derived from the LAB survey by
\citet{Kalberla2005}. For details on the analysis see \citet{hiemiss}.

\begin{figure}[!tb]
 \includegraphics[width=0.8\textwidth]{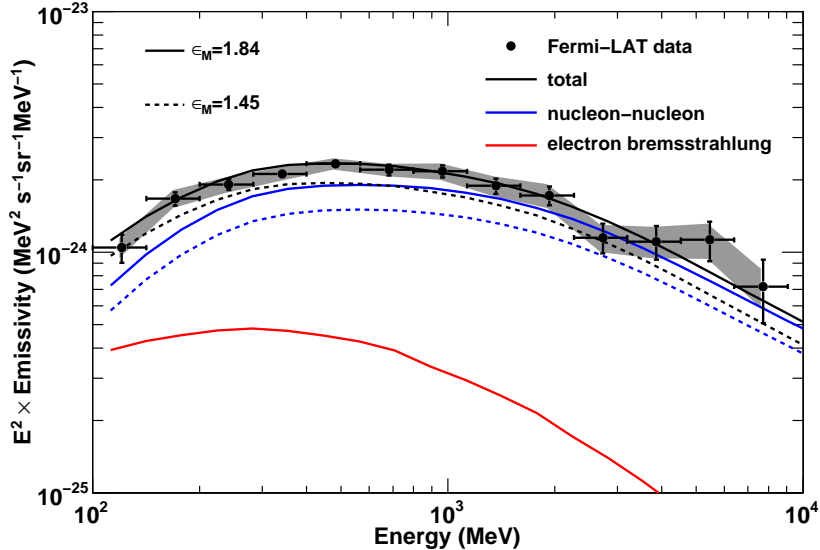}
 \caption{\citep{hiemiss} Spectrum of $\hi$ emissivity measured by the LAT
(points) compared with predictions based on the locally measured CR
spectra. Black lines are the total $\hi$ emissivity, other lines are the 
$\pi^0$-decay component and the electron Bremsstrahlung component.
Dashed and solid lines correspond to
different values of the nuclear enhancement factor
(see text).}\label{hiemissfig}
\end{figure}

The obtained spectrum of $\hi$ emissivity is shown in Fig.~\ref{hiemissfig}.
The emissivity measured by the LAT is compared with the results of a
calculation based on
the locally measured CR spectra. The Bremsstrahlung
component is derived using GALPROP. In the energy range considered the dominant
contribution is the $\pi^0$-decay emission. Using the locally measured proton
spectrum we calculated the $p$-H \g-ray emissivity using the inclusive
parametrization by \citet{Kamae2006}. The obtained emissivity spectrum is then
multiplied by a factor $\varepsilon_\mathrm{M}$ which accounts for interactions
involving heavier nuclei
in both CRs and the ISM. This factor, often called nuclear enhancement
factor, is the subject of a long-standing debate in the literature, where
values spanning from 1.45 \citep{Dermer1986} to 1.84 \citep{Mori2009} are
found. The two extreme values are considered in Fig.~\ref{hiemissfig} for
comparison. The results show that \g-ray emission from local gas is consistent
with CR spectra measured at Earth.

\section{The distribution of CR sources and the $\protect{\mathbf{\xco}}$
ratio in the outer Galaxy}
Since the local interstellar \g-ray emission is
consistent with models based on locally measured CR spectra, we
are confident that we have understood its basic mechanisms, and, as explained
in the introduction, we can use diffuse emission as a tracer of CR densities and
of ISM column densities throughout the Galaxy. The first target was the outer
Galaxy, in selected regions where the velocity gradient with galactocentric
radius is very steep, and so the Doppler shift of the radio/microwave lines
observed for interstellar gas
provides a good separation of the different structures along each line of sight.

As told above, the 21 cm line of $\hi$ allows us to directly trace its
column densities. Molecular hydrogen, $\hd$, which is the most abundant
constituent of the molecular phase of the ISM, does not have
observable lines. The integrated intensity $\wco$ of the 2.6 mm line of CO is
usually adopted as a surrogate tracer of molecular masses, assuming that $\nhd=
\xco \cdot \wco$. However, the $\xco$ ratio is still uncertain,
and in particular there are many evidences that it might increase in the outer
Galaxy with respect to local clouds, as explained in \citet{Strong2004grad}
and references therein. The increase still needs to be verified in \g-rays,
because of the limited performances of previous-generation telescopes
\citep{Digel1996,Digel2000}.

On the other hand for many years supernova remnants (SNRs) have been considered
the best candidates as CR sources in the Galaxy. However, the origin of CRs is
still mysterious and the distribution of SNRs is very poorly determined
\citep{Case1998}, leading to large uncertainties in the models of diffuse
\g-ray emission.

Since the Doppler shift of the radio/microwave lines allows us the to separate
different
structures along a line of sight these issues can be investigated directly in
\g-rays:
\begin{itemize}
 \item the emissivity of the diffuse $\hi$ gas can be used to trace the CR
densities in distant locations;
 \item the emissivity per $\wco$ unit, compared with that per $\hi$ atom, can be
used to evaluate the $\xco$ ratio in each molecular complex.
\end{itemize}

\begin{figure}[!bht]
\includegraphics[width=0.8\textwidth]{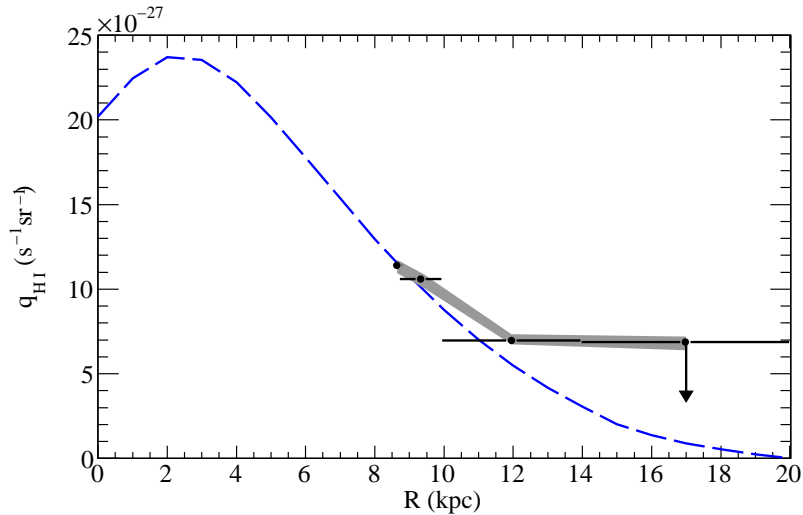}
\caption{\citep{cascep} Emissivity of atomic gas integrated above
200 MeV, $q_\mathrm{H\,\scriptscriptstyle{I}}$, as a function of Galactocentric
radius: points are LAT measurements (the shaded area represents systematic
uncertainties in the event selection efficiency), the dashed line is the
prediction
by a GALPROP model based on a CR source distribution derived from
pulsars \citep{Lorimer2004}.}\label{cascepfig1}
\end{figure}

\begin{figure}[!thb]
\includegraphics[width=0.8\textwidth]{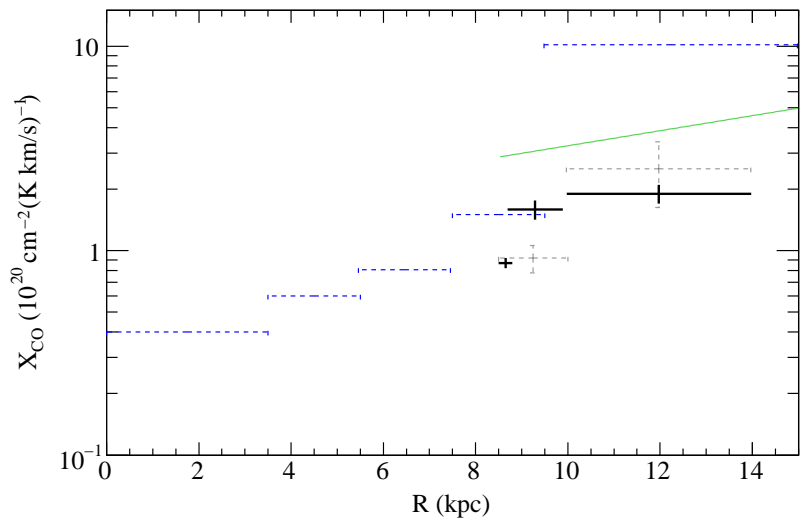}
\caption{\citep{cascep} $\xco$ as a function of
Galactocentric radius: black points are LAT measurements in
the second quadrant, gray points EGRET measurements in the
same region. The dashed step function represents the model used in
GALPROP by \citet{Strong2004grad} and the solid line is the
conversion law based on virial masses by
\citet{Nakanishi2006}.}\label{cascepfig2}
\end{figure}

The first region studied with this method was the region of Cassiopeia and
Cepheus in the second Galactic quadrant. For details on the analysis see
\citet{cascep}. The main results are shown in
Fig.~\ref{cascepfig1}~and~\ref{cascepfig2}.

Fig.~\ref{cascepfig1} shows that the gradient of $\hi$
emissivity measured
by the LAT is flatter than the predictions by a GALPROP model based on a CR
source distribution peaking in the inner Galaxy as suggested by pulsars. This
might point to CR sources extending further in the outer Galaxy, but might also
hint to
diffusion parameters different from those derived from
local CR isotopic abundances used by GALPROP. However, systematic
effects might be at the origin of this effect, e.g. contamination by
populations of unresolved sources or large errors in the determination of
$\nhi$ (because of the approximations applied to handle the radiative transfer
of $\hi$ radio emission
or self-absorption phenomena).

Fig.~\ref{cascepfig2} confirms a significant but moderate increase of $\xco$
from the solar circle to
the outer Galaxy. The large gradient proposed by \citet{Strong2004grad} to
interpret EGRET data is not supported by our results. This large gradient was
introduced on the basis of non-\g-ray data to accommodate the problem of the
flat emissivity profile, but this solution is disfavored by the LAT.
On the other hand the values are systematically lower
than the conversion law by \citet{Nakanishi2006}, mostly based on virial
masses. The \g-ray estimates might be biased by the limited resolution of the
data, but on the other side the virial masses are based on the rather crude
assumption of clouds with simple spherical shapes and a velocity dispersion due
only to balancing gravitational forces. 

Further investigation is undergoing by the LAT team to extend the analysis to
the segment of the outer Galaxy seen in the third quadrant and to explore the
origin of the flat gradient of $\hi$ emissivities \citep{3quad}.
 
\section{The Large Magellanic Cloud}
External galaxies are a promising target for diffuse emission studies,
because the interpretation of the data is less
affected by confusion along the line of sight than in the Milky Way. The only
normal galaxy seen in high-energy \g-rays before
the LAT era has been the Large Magellanic Cloud (LMC) \citep{Sreekumar1992}. It
is an ideal object to map CR acceleration sites and study their propagation in
the ISM because it is nearby ($\sim 50$ kpc), seen at a small inclination angle
($20^\circ-35^\circ$) and it hosts many SNRs, bubbles and star-forming
regions. EGRET, due to its limited angular resolution, could not resolve the
LMC,
but the measured integral flux led to derive CR densities similar to those in
the Milky Way. On the other hand, the non detection by EGRET of the Small
Magellanic Cloud (SMC), a
similar external Galaxy, was a strong evidence supporting the idea that CRs up
to energies of $\sim 10^{15}$ eV are galactic and not extragalactic in origin
\citep{Lin1996}.

The LAT, thanks to its improved angular resolution, is now able to resolve the
LMC, making it the first extragalactic object ever resolved in high-energy
\g-rays.
\begin{figure}[!bhtp]
 \includegraphics[width=0.75\textwidth]{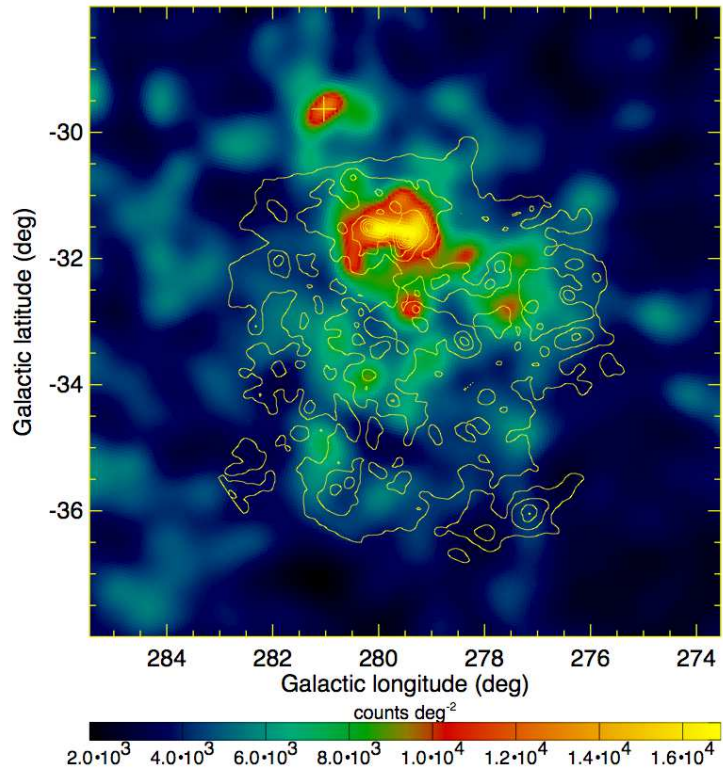}
 \caption{\citep{PorterLMC} Adaptively smoothed preliminary map of LAT counts
(200 MeV -- 100 GeV, from the first $~200$ days of the Science phase of the
mission) in the region of the Large Magellanic Cloud. The starbust region 30
Doradus appears at $l\sim279.5^\circ$ $b\sim-31.5^\circ$. Overlaid contours of
ionized gas density in the LMC \citep{Finkbeiner2003}. 
The cross in the north-east of the image corresponds to
the location of the blazar CRATES J060106-703606.}\label{lmcfig}
\end{figure}

In Fig.~\ref{lmcfig} we show a preliminary count map of the LMC. The starbust
region of 30 Doradus
is clearly visible together with a larger halo of diffuse emission,
plausibly arising from interactions of CRs with the gas in the ISM. A detailed
analysis of
this object, investigating CR acceleration and propagation in the LMC, will be
presented soon by the LAT team \citep{LMC}.

\section{Summary and perspectives}
In the previous sections we have presented some results relevant for
the understanding of diffuse \g-ray emission, obtained by the \textit{Fermi}
LAT during the first year of observations: the non-confirmation of the
EGRET GeV excess in the local interstellar emission, the measurement of the
\g-ray emissivity of local atomic gas, the study of the CR density and
$\xco$ gradient across the Galaxy and the first view of diffuse emission in an
external Galaxy.

All these findings are being incorporated in a new large-scale model of diffuse
\g-ray emission based on the previously introduced GALPROP code \citep{GDE}. As
we have seen, \emph{a priori} GALPROP predictions are locally
approximately consistent with LAT measurements of interstellar \g-ray emission,
so reasonable adjustments of the CR spectra, CR source distribution, CR
propagation parameters (e.g. the height of the propagation halo, the diffusion
coefficients), ISM properties
(e.g. $\xco$) are expected to provide a large-scale agreement
between observations and physical expectations \citep{GDE}. At the
same time the
comparison will help to get a deeper understanding of the long-standing
puzzle of CR acceleration and propagation, as well as about the census of the
gas in the ISM.

The Galactic interstellar emission model will support a new estimate of
the isotropic diffuse spectrum, and later on of the EGB spatial properties,
based on LAT data. The new estimate of the isotropic diffuse spectrum is based
on event selection criteria explicitly developed for the delicate task of
separating the EGB from residual backgrounds due to CR interactions in the
LAT misclassified as \g-rays. The isotropic spectrum is then derived from a
fit to LAT data including the aforementioned model of Galactic interstellar
emission, LAT resolved sources and the emission from the Sun. The analysis
and results will be presented in detail in \citet{EGB}. These studies,
together with source population syntheses, will provide insightful information
on the nature of the EGB.

As LAT data accumulate we can expect many more results, which will make the
next years very exciting for studies on diffuse \g-ray emission.


\begin{theacknowledgments}
The \textit{Fermi} LAT Collaboration acknowledges support from a number of
agencies and institutes for both development and the operation of the LAT as
well as scientific data analysis. These include NASA and DOE in the United
States, CEA/Irfu and IN2P3/CNRS in France, ASI and INFN in Italy, MEXT, KEK, and
JAXA in Japan, and the K.~A.~Wallenberg Foundation, the Swedish Research Council
and the National Space Board in Sweden. Additional support from INAF in Italy
and CNES in France for science analysis during the operations phase is also
gratefully acknowledged.

LT is partially supported by the International Doctorate on AstroParticle
Physics (IDAPP) program.
\end{theacknowledgments}

\bibliographystyle{aipproc}

\end{document}